# Self-sensitive torsional microresonators based on a charge-density wave system

V.Ya. Pokrovskii & S.G. Zybtsev

*Institute of Radioengineering and Electronics of RAS, Mokhovaya 11-7, 125009 Moscow, RUSSIA.*

**Recently, there have been dramatic advances in the miniaturization of electromechanical devices. Most of the micro- and nanoelectromechanical systems (MEMS-NEMS) operate in the resonant modes[1]. The micron-, and, the more, the submicron-sizes devices, are driven usually by electrostatic forces, as piezoelectric and electromagnetic engines are powerless over this size range. Such engines could play the role of external actuators for the NEMS, being, however, macro devices in their own. Apart from actuation, an objective of NEMS is getting the output signal characterizing the oscillations[1], so, actuators sensing their own motion are rather topical (see [2] for example). Not long ago, several works appeared demonstrating high sensitivity of the sizes[3,4] and form[5] of quasi one-dimensional conductors to the deformations of the charge-density wave (CDW). Here we demonstrate electrically driven torsional resonators based on whiskers of the quasi one-dimensional conductor with CDW, $TaS_3$. The driving force for the torsional deformation is peculiar to the CDW systems and reflects the transmission of the CDW deformation to the crystalline lattice. In comparison with the piezoelectrics, the effect of electric field on the crystal deformation is 3-4 orders of magnitude larger. The resonator is found to provide also a torsion-induced electrical feed-back (output signal) from the oscillations. We discuss the CDW systems as promising elements for NEMS-MEMS.**

The CDW deformation under electric field is well studied in itself [6,7], however its mechanical effects on the crystal are in the early stage of studies and understanding. The CDW forms in quasi one-dimensional conductors below the Peierls transition temperature $T_P$, which is typically hundreds of Kelvin, and for some compounds can approach and even exceed the room temperature[6,7]. The CDW is a 3-dimensional electronic crystal inside the pristine lattice. Strongly coupled to the electric field $E$, it can deform while $E$ is below the threshold field $E_t$, depin (dart off) at $E_t$ and slide at $E > E_t$. The CDW deformations give rise to metastable states, which reveal themselves in different properties[7], including the dimensions[3,4] and form[5]. The coupling of the CDW deformation to the lattice – revealing itself also in strong influence of electric field on the Young and shear moduli of the crystals[8] – still has no unambiguous explanation.

Recently it has been found that the CDW depinning is accompanied by torsional strain of the whiskers[5]. The features of the strain are threshold behavior and hysteresis *vs*. $E$; its observation has revealed surface pinning and a polar axis in the CDW state. With shear achieving $10^{-4}$ at the surface under several hundred mV/cm, one can attribute an enormous "piezomodulus" to the effect – above $10^{-4}$ cm/V. The torsional strain has been observed[5] for $TaS_3$, a typical CDW conductor from the chalcogenide group[6] with $T_P=220$ K. $TaS_3$ grows as needle-like crystals (whiskers) up to several mm long and up to several tens of μm wide. Below $T_P$, $TaS_3$ from a metal turns into a semiconductor whose resistance grows with the activation energy about 800 K, corresponding with the Peierls gap value, and in the range of liquid-nitrogen temperatures resistivity of $TaS_3$ is of the order of 0.1-1 Ωcm..

To enable torsional strain, one should suspend one of the sample contacts. For the observation of the oscillations we developed three variants of sample arrangement (Fig. 1). In the variant **a)** the electric contact to the elevated end is supplied by a thin low-Ohmic wire[5] – here a HTSC whisker of $Ba_2Sr_2CaCu_2O$ (BSCCO). The mirror(s)

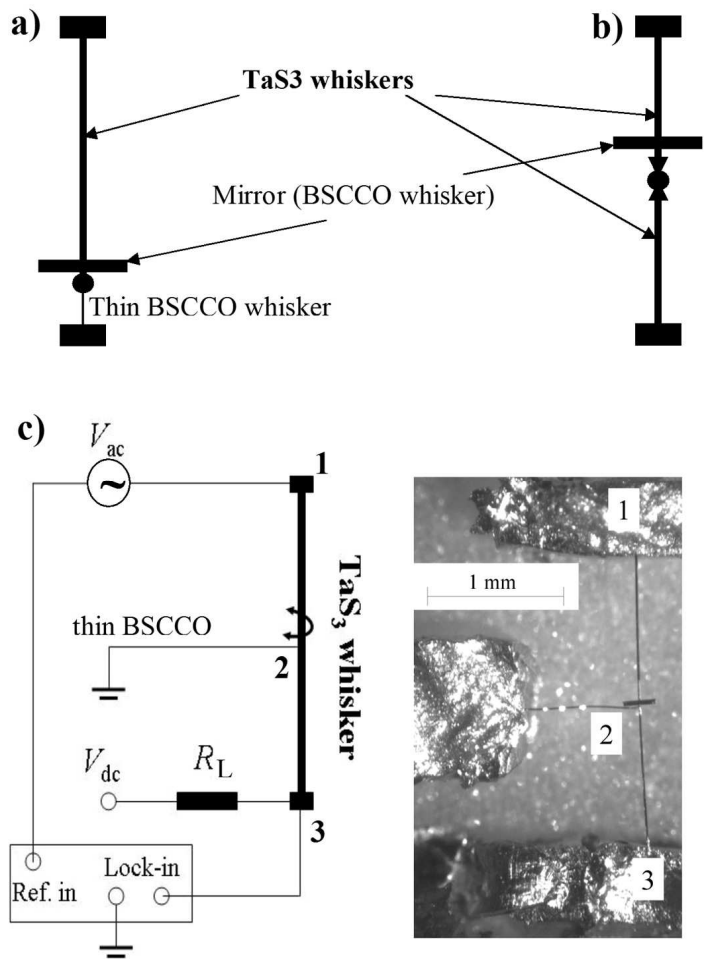

**Figure 1** Schemes of samples arrangements for observation of torsional oscillations. In **b)** the arrows point out the opposite directions of the polar axes. The scheme **c)** is supplied with the electric scheme of the actuation and the feedback of the oscillations, and with a microphotograph of one of the samples. A micromirror is attached slightly above the contact with the BSCCO whisker.

(also, BSCCO whiskers) stuck to the sample enable optical registration of the torsional strain basing on the turning of the reflected laser beam. Variant **b)** exploits the orientation of the polar axis[5]: the two samples connected in the center are parts of one whisker cut crosswise. One of the pieces is turned upside down, so that the *c*-axes, parallel to the longer dimension, are oriented oppositely. Electric field $\gtrsim E_t$ forces the adjacent (the soldered) whiskers ends to turn in the same direction[5]. No additional wire is needed, so the parasitic elasticity is completely excluded. The highest $Q$ has been achieved in this arrangement. Variant **c),** the central one for this Letter, has allowed obtaining a feedback signal from the resonator. Both ends (contacts 1 and 3) of the suspended $TaS_3$ whisker are fixed, while a flexible thin whisker (BSCCO or $NbSe_3$) forms an additional contact (No 2) near the center. Voltage 1-2, $V_{12}$, results in the torsional strain of the upper part of the whisker turning the sample center around the *c*-axis[5]. The strain is transmitted to the lower part. Measuring the 2-3 resistance we observe its torsional modulation. Thus, the upper part of the whisker (1-2) works as an actuator, and the lower (2-3) – as a receiver (or *vice versa*). The mirror here is a "rudiment", in a sense, as in this configuration one does not need peeping at the torsional oscillations through the cryostat window.

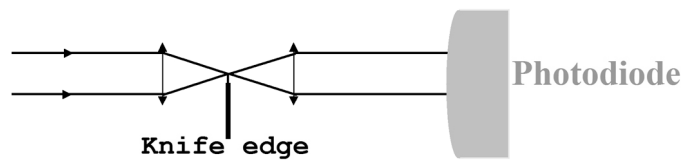

**Figure 2** The optical scheme for the detection of beam deflection with a photodiode.

The technique proposed by A.P. Kuznetsov[9] allows to transform the deflection of the beam reflected from the mirror directly into intensity: the beam passes through a collimator with a knife edge in the focus (Fig. 2) and then falls on the photodiode. Initially, the edge is shutting about half the beam. With the beam deflection, the spot in the focus moves, and the output intensity changes. This very sensitive method allows registration of the high-frequency (>50 kHz at least) torsional oscillations.

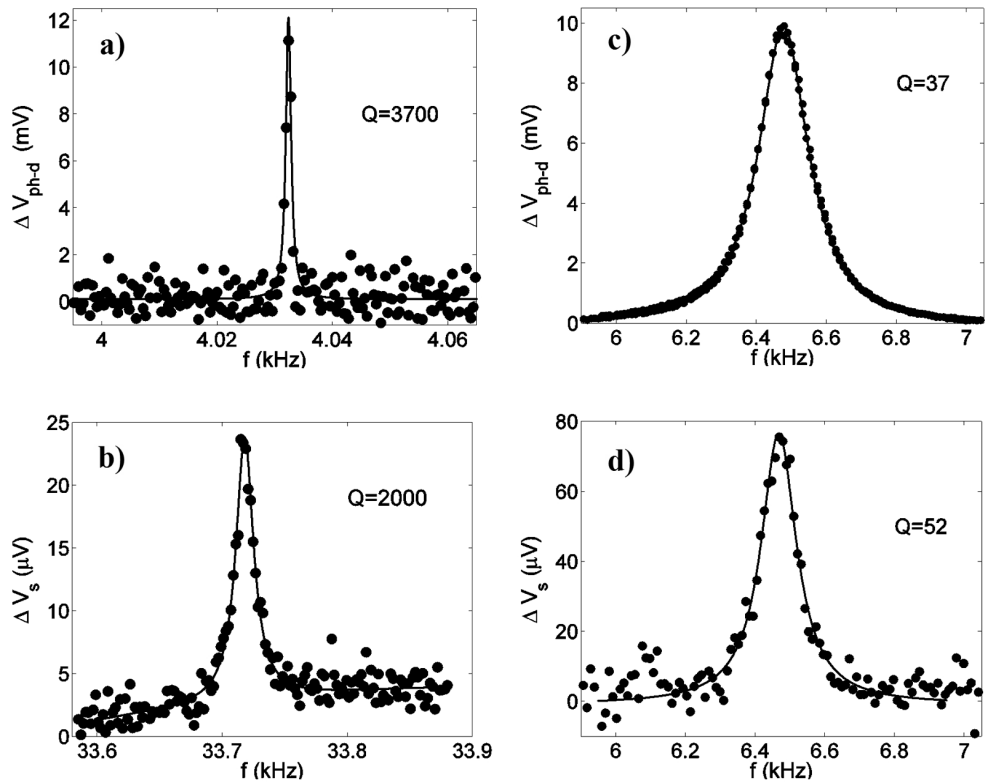

**Figure 3** Frequency dependences of the oscillations detected with the lock-in techniques, indicating the resonance peaks. The solid lines are Lorentzian fits. **a.** signal from the photodiode (configuration Fig. 1b). The sample is composed of two pieces 0.87 and 0.95 mm long, cross-section area is about 2μm$^2$, $T$=79 K. **b.** detected AC voltage across the contacts 3-2 (configuration Fig. 1c). The torsional resonance at 33.72 kHz is a higher natural mode (the 1$^{st}$ resonance is observed at 3.3 kHz). The 15 μV peak (rms.) at $I$=13 μA ($I_t$≈1 μA), corresponds to $\Delta R/R$ modulation about 5·10$^{-5}$ under actuating AC voltage, $V_{12}$= ±300 mV. The sample length is 3 mm, cross-section area is about 10μm$^2$, $T$=79 K. **c,d:** configuration Fig. 1c. Signal from the photodiode (**c**) and the AC voltage from contacts 3-2 (**d**) under identical conditions: $T$= 58 K, actuating AC voltage ±1.05 V. The 70 μV peak (rms) at DC current $I$=175 μA ($I_t$≈35 μA) corresponds to $\Delta R/R$ modulation about 10$^{-4}$. The sample length is 3 mm, cross-section area ~ 100 μm$^2$. The low $Q$-factors are due to the damping effects of the high current[8] and the transverse wire (Fig. 1c).

Figure 3a shows the frequency dependence of the signal from the photodiode for the sample in the 2$^{nd}$ arrangement (Fig. 1b), detected by a lock-in tuned in phase at the resonance frequency. One can see a narrow maximum at $f_0$=4.03 kHz. Here the AC voltage is below the threshold value, so that the amplitude at the maximum is about 0.2$^o$. For higher voltages the $Q$-factor drops[8]. $Q$ and $f_0$ grow notably with lowering temperature, the highest values of $Q$ exceeding 10$^4$. The resonant frequencies observed fall in the range 1–15 kHz and are lower than the value estimated from the formula for a uniform cylinder[10] $f_0$=(1/2$l$)√($G$/ρ)  (≈ 200 kHz  with $G$ =5 GPa, the shear modulus[8], ρ=8.5*10$^3$ kg/m$^3$, the density of TaS$_3$[6], and $l$=1.8 mm, the sample length – Fig. 3b). This, presumably, should be assigned to the inertia moment of the mirrors.

Figures 3b and d show the signal directly from the TaS$_3$ samples in the configuration shown in Fig. 1c. A constant current through contacts 3-2, $I_{32}$, is provided by a DC voltage source and the high loading resistance, $R_L$. The modulated (AC) component of the voltage across the sector 3-2, $V_{32}$, detected synchronously with the AC voltage by a lock-in is shown in Figures 3b and d. A clear maximum of $V_{32}$ is detected at the resonant frequency 33.72 kHz (this is a higher natural torsional mode). Figures 3c and 3d show the signals from another sample in the same configuration, detected by the photodiode and as the resistance modulation under identical conditions. Both methods reveal a torsional resonance at the same frequency (somewhat lower $Q$ in Fig. 3c might be attributed to the nonlinearity of the detection technique (Fig. 2) for high amplitudes). Thus, the TaS$_3$ whisker works here as a torsional piezoresistor (a "torsistor").  For all the samples the peak height (*i.e.* the resistance modulation, Δ$R$/$R$) appeared to be of the same order as the shear deformation at the surface, Δφ$w$/$l$, typically ~10$^{-4}$.  Thus we can estimate the ratio between Δ$R$/$R$ and the shear deformation – the torsio-resistivity – to be of the order of unity. Up to our knowledge, this is the 1$^{st}$ indication of the effect. The peaks in $V_{32}$ (Figures 3b and 3d) are observed at forevacuum, while at ambient pressure of helium gas only a hardly discernible feature remains. They are observed only under current exceeding the threshold, $I_t$, the best signal-to-noise ratio is achieved typically at $10I_t$. The strain-induced voltage needs a further detailed study.

The result in Figure 3b is a demonstration of a self-sensitive resonator. In contrast to the conventional micro-devices driven by electrostatic force, the present resonator does not need any external agent, like a gate-like electrode: the whisker comprises the engine inside itself. The closest parallel is piezoelectric effect. In terms of piezomoduli, the CDW-based actuators are at least 3-4 orders of magnitude "stronger"[5]. However, the difference from the piezoelectric effect is not only quantitative: the torsional strain is fundamentally non-uniform, the shear growing from the center to the surfaces. Meanwhile, it develops under the conditions of uniform electric field. Here the non-uniform nature of deformation reveals the non-local properties of CDW[7]: evidently, the surface pinning impels high CDW deformation at the surface. For illustration, one can confront our devices with a piezoelectric torsional resonator[10], which has required special arrangement of piezoceramic pieces with different polarization directions. Under voltage ±800 V the actuator[10] 25 mm long supplied with a torsion bar 126 mm long showed a maximum angular displacement 0.18° in the resonant mode.

For the present demonstration we did not make special efforts for miniaturization of the devices. However, we do not see principal complications in reducing the length below 1 μm, and the thickness below 0.1 μm. In principle, such devices could be fabricated even manually[11], without involving special nanotechnology. At such dimensions the whiskers retain their CDW properties, though the fields to be applied can grow appreciably[11,12]. Note, that the thinner is a whisker, the higher is the torsional angle[5].

Thus, we have demonstrated an actuator based on the CDW system, driven by it and providing an output signal. Driven like a piezoelectric, it is low-Ohmic enough to provide a measurable tensoresistive-like response. Up to our best knowledge, such resonators have no analogues. Clearly, the prospects of their optimization are wide. Apart from miniaturization, one can improve the *Q*-factor by perfection of clamping. The working current, voltage and temperature ranges, the feedback circuitry also require optimization. One should also try to search for the feedback in the 2-probe configuration, like it is the case in the quartz generators. A promising mission is to test other CDW compounds, among which one can find chalcogenides with high $T_P$ – close to the room temperature or above it: $NbS_3$, ($T_P$=340 K), $(TaSe_4)_2I$ (263 K), $(NbSe_4)_{10}I_3$ (285K), $(NbSe_4)_3I$ (274 K)[6]. It is early to specify, what particular MEMS-NEMS devices will be based on the CDW actuators, though, the general challenges of this scientific and technical area can by addressed to the devices proposed here. By the way, the torsional modes of oscillations do not require such high vacuum, as the flexural ones. We can note, that in the present state the resonator is a nearly ready self-sensitive cantilever (for an AFM, *e.g.*), whose advantage is the growth of sensitivity with lowering temperature.

**Acknowledgements** We are grateful to S.V. Zaitsev-Zotov, J.W. Brill and A.P. Kuznetsov for useful discussions and to R.E. Thorne for the high-quality samples. The support of ISTC, RFBR, RAS programs

``New materials and structures'' and of the RAS Presidium is acknowledged. The research was held in the framework of the CNRS-RAS-RFBR Associated European Laboratory ``Physical properties of coherent electronic states in condensed matter'' including CRTBT and IRE.

**Correspondence** and requests for materials should be addressed to V.Ya.P. (pok@cplire.ru).